\documentclass{paper}

\usepackage{epsfig}
\usepackage{graphicx}

\begin{document}
\title{Black Holes in Two Dimensions}
\author{Yuri N. Obukhov and Friedrich W. Hehl}

\institution{Dept. of Theoretical Physics,
Moscow State University, 117234 Moscow, Russia and
Inst. for Theoretical Physics,
University of Cologne, D-50923 K\"oln, Germany}

\maketitle

\begin{abstract}
  Models of black holes in $(1+1)$-dimensions provide a theoretical
  laboratory for the study of semi-classical effects of realistic black
  holes in Einstein's theory. Important examples of two-dimensional
  models are given by string theory motivated {\em dilaton
    gravity}, by ordinary general relativity in the case of {\em
    spherical symmetry}, and by {\em Poincar\'e gauge gravity} in two
  spacetime dimensions.  In this paper, we present an introductory
  overview of the exact solutions of two-dimensional classical
  Poincar\'e gauge gravity (PGG).  A general method is described with
  the help of which the gravitational field equations are solved for
  an arbitrary Lagrangian. The specific choice of a torsion-related
  coframe plays a central role in this approach.  Complete
  integrability of the {\it general} PGG model is demonstrated in
  vacuum, and the structure of the black hole type solutions of the
  {\it quadratic} models with and without matter is analyzed in
  detail. Finally, the integrability of the general dilaton gravity
  model is established by recasting it into an effective PGG model.
  {\em file tworev8.tex, 1998-07-14}
\end{abstract}

\section{Introduction}\index{dilaton gravity}\index{PGG (Poincar'e 
gauge gravity)}

Standard General Relativity (GR) is trivial in two dimensions.
Nevertheless, two--dimensional (2D) models of gravity which differ
from GR have recently received considerable attention
\cite{akd1}-\cite{russo2}.  The interest in 2D gravity is strongly
supported by the fact that usual four-dimensional GR, in the case of
spherical symmetry, is described by an effective 2D gravitational
model of the dilaton type. Such a dimensional reduction provides a
technical tool for the study of long standing problems in black hole
physics, including an understanding of the final state of a black
hole with an account of the back reaction of the quantum evaporation
process, see \cite{cal,banks,russo2}. On the other hand,
lower--dimensional black hole physics is discussed in the context of
string theory motivated dilaton gravity (see, e.g.,
\cite{cal,kat6,str2,str3,kumfermi,nappi1,nappi2,nappi3,russo}) and in the
framework of PGG (``Poincar\'e gauge gravity'')
\cite{ana}-\cite{can3},\cite{o1,o2,o4},\cite{sol2}-\cite{sol6}. The
approaches \cite{ana}-\cite{can3} are attempts to construct string
theories with a dynamical geometry \cite{kat1}-\cite{kat8},\cite{o4}.
In this paper we present an overview of the black hole solutions in
classical two--dimensional PGG. At the same time, 2D gravity is of
interest in itself as a theoretical laboratory which offers a simple
way to study difficult non-perturbative quantization problems
\cite{kat4}.

In the studies of both, classical and quantized 2D models, it is of
crucial importance to find exact solutions of the field equations.
Here we describe an elegant method developed in \cite{o1}-\cite{sol6}
with the help of which one can explicitly integrate the field
equations of classical PGG with and without matter sources. The
central point is to use a specific coframe built up from the one--form
of the torsion trace and its Hodge dual. The early proofs of the
integrability of the {\it quadratic} PGG models in vacuum were based
on the component approach and relied on specific gauge choices, like
the conformal or the light-cone gauge \cite{kat1}-\cite{kum6}. The
coupling to gauge, scalar, and spinor matter fields was shown to
destroy the integrability in general. A peculiar but common feature of
the {\em standard} matter sources (gauge, scalar, and spinor fields)
in two dimensions is that for all of them the spin current vanishes.
Thus, quite generally, the Lorentz connection is explicitly decoupled
from 2D matter. Hence the material energy--momentum current is
symmetric and covariantly conserved with respect to the Riemannian
connection. The absence of the spin--connection coupling considerably
facilitates the integration of the field equations.

\begin{figure}[ht]
  \begin{center}
    \leavevmode
\begin{picture}(0,0)%
\epsfig{file=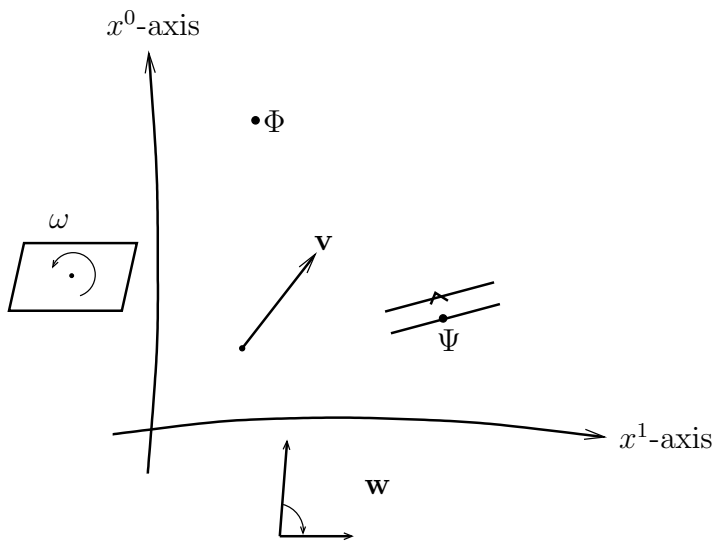}%
\end{picture}%
\setlength{\unitlength}{3947sp}%
\begingroup\makeatletter\ifx\SetFigFont\undefined%
\gdef\SetFigFont#1#2#3#4#5{%
  \reset@font\fontsize{#1}{#2pt}%
  \fontfamily{#3}\fontseries{#4}\fontshape{#5}%
  \selectfont}%
\fi\endgroup%
\begin{picture}(3872,3337)(159,-2558)
\put(2886,-1371){\makebox(0,0)[lb]{\smash{\SetFigFont{12}{14.4}{\familydefault}{\mddefault}{\updefault}$\Psi$}}}
\put(1796,  4){\makebox(0,0)[lb]{\smash{\SetFigFont{12}{14.4}{\familydefault}{\mddefault}{\updefault}$\Phi$}}}
\put(451,-601){\makebox(0,0)[lb]{\smash{\SetFigFont{12}{14.4}{\familydefault}{\mddefault}{\updefault}$\omega$}}}
\put(2431,-2266){\makebox(0,0)[lb]{\smash{\SetFigFont{12}{14.4}{\familydefault}{\mddefault}{\updefault}{\bf w}}}}
\put(811,614){\makebox(0,0)[lb]{\smash{\SetFigFont{12}{14.4}{\familydefault}{\mddefault}{\updefault}$x^0$-axis}}}
\put(4031,-1981){\makebox(0,0)[lb]{\smash{\SetFigFont{12}{14.4}{\familydefault}{\mddefault}{\updefault}$x^1$-axis}}}
\put(2116,-736){\makebox(0,0)[lb]{\smash{\SetFigFont{12}{14.4}{\familydefault}{\mddefault}{\updefault}{\bf v}}}}
\end{picture}
    \caption{
\emph{Two-dimensional spacetime}:
A 0-form (scalar) $\Phi$ has one component, a 1-form $\Psi$ two components,
and a 2-form $\omega$ one component.
A vector \textbf{v} has two components, a bivector \textbf{w} one component.
}
  \end{center}
\end{figure}

\begin{figure}[htb]
  \begin{center}
    \leavevmode
\begin{picture}(0,0)%
\epsfig{file=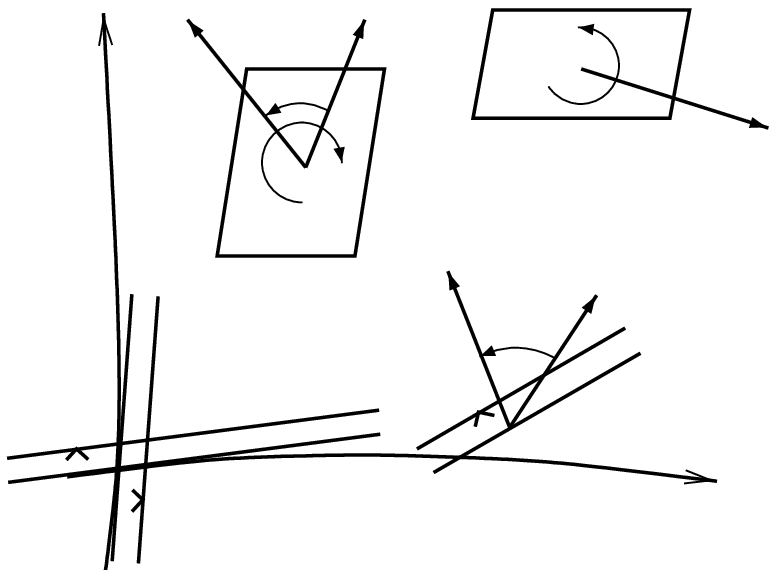}%
\end{picture}%
\setlength{\unitlength}{3947sp}%
\begingroup\makeatletter\ifx\SetFigFont\undefined%
\gdef\SetFigFont#1#2#3#4#5{%
  \reset@font\fontsize{#1}{#2pt}%
  \fontfamily{#3}\fontseries{#4}\fontshape{#5}%
  \selectfont}%
\fi\endgroup%
\begin{picture}(3595,3152)(501,-2373)
\put(811,614){\makebox(0,0)[lb]{\smash{\SetFigFont{12}{14.4}{\familydefault}{\mddefault}{\updefault}$x^0$-axis}}}
\put(4031,-1981){\makebox(0,0)[lb]{\smash{\SetFigFont{12}{14.4}{\familydefault}{\mddefault}{\updefault}$x^1$-axis}}}
\put(1736,109){\makebox(0,0)[lb]{\smash{\SetFigFont{12}{14.4}{\rmdefault}{\mddefault}{\updefault}$R^{\alpha\beta}$}}}
\put(4096,-466){\makebox(0,0)[lb]{\smash{\SetFigFont{12}{14.4}{\rmdefault}{\mddefault}{\updefault}$T^\alpha$}}}
\put(501,-1681){\makebox(0,0)[lb]{\smash{\SetFigFont{12}{14.4}{\rmdefault}{\mddefault}{\updefault}$\vartheta^{\hat 0}$}}}
\put(1286,-2291){\makebox(0,0)[lb]{\smash{\SetFigFont{12}{14.4}{\rmdefault}{\mddefault}{\updefault}$\vartheta^{\hat 1}$}}}
\put(3101,-1766){\makebox(0,0)[lb]{\smash{\SetFigFont{12}{14.4}{\rmdefault}{\mddefault}{\updefault}$\Gamma^{\alpha\beta}$}}}
\end{picture}
    \caption{
\emph{Two-dimensional Riemann-Cartan spacetime}:
Coframe $\vartheta^\alpha$ (2 components),
connection $\Gamma^{\alpha\beta}=-\Gamma^{\beta \alpha}$ (2 components),
torsion $T^\alpha$ (2 components),
curvature $R^{\alpha \beta}=-R^{\beta \alpha}$ (1 component).
Here, $\vartheta^\alpha=\{\vartheta^{\hat 0}, \vartheta^{\hat 1}\}$ is a
natural coframe, i.e.,  $\vartheta^{\hat 0}=d x^0$, $\vartheta^{\hat 1}=d x^1$.
}
  \end{center}
\end{figure}

The structure of the paper is as follows: Sec.\ \ref{obh:sect2} contains
an introduction to 2D Riemann-Cartan geometry. In Secs.\ \ref{obh:sect3}
and \ref{obh:sect4}, we demonstrate the integrability of PGG with an
arbitrary gravitational Lagrangian and prove the consistency of our
method in general. As a particular application, a {\it quadratic}
model with an action containing squares of torsion and curvature is
discussed in vacuum (Sec.\ \ref{obh:sect5}) and in the presence of
conformally invariant matter (Sec.~\ref{obh:sect6}). The properties of the
exact solutions of black hole type are described in detail.  Finally,
in Sec.~\ref{obh:sect7}, we apply the general method to the (purely {\it
  Riemannian}) string theory motivated dilaton gravity models by
rewriting them in form of an effective {\it Poincar\'e-Brans-Dicke}
theory. The general solution of an arbitrary two-dimensional dilation
gravity model is obtained in explicit form.

\section{Two-dimensional Riemann-Cartan 
space\-time}\label{obh:sect2}\index{Riemann-Cartan spacetime}

The Riemann--Cartan geometry has rather remarkable properties in two 
dimensions, see Fig.1. 

In the PGG approach, the {\em orthonormal} coframe one--form
$\vartheta^{\alpha}$ and the linear connection one-form
$\Gamma^{\alpha\beta}$ are considered to be the translational and the
Lorentz gauge potentials of the gravitational field, respectively. The
corresponding field strengths are given by the torsion
two-form\index{torsion 2-form} $T^{\alpha} := D\vartheta^{\alpha}$ and
the curvature two-form $R^{\alpha\beta}:= d\Gamma^{\alpha\beta}-
\Gamma^{\alpha\gamma}\wedge\Gamma_{\gamma}{}^{\beta}$, see Fig.2. The
frame $e_{\alpha} = e^{i}{}_{\alpha}\,\partial_{i}$ is dual to the
coframe $\vartheta^{\beta} = e_{j}{}^{\beta}\,dx^{j}$, i.e.,
$e_{\alpha}\rfloor\vartheta^{\beta} =
e^{i}{}_{\alpha}\,e_{i}{}^{\beta} = \delta_{\alpha}^{\beta}$. The
spacetime manifold $M$ is equipped with a metric
\begin{equation}
g = g_{ij}\,dx^{i}\otimes dx^{j}\, .
\end{equation}
Thus its coframe components satisfy
\begin{equation}
o_{\alpha\beta} = e^{i}{}_{\alpha}\,e^{j}{}_{\beta}\,g_{ij}\; ,
\qquad\qquad (o_{\alpha\beta}) ={\rm diag}(-1,+1).  
\end{equation}
In an orthonormal frame, the curvature, like the connection, 
is antisymmetric in $\alpha$ and $\beta$. 

\begin{table}[ht]\label{obh:table1}
\centering
\caption{Gauge field strengths, matter currents, and $\eta$--basis}
\begin{tabular}{|c|c|c|c|c|}
\hline\hline
&\multicolumn{2}{|c|}{\it Type}&\multicolumn{2}{c|}{\it Components}\\
\hline
{\it Object}&{\it Valuedness}&{\it p--form}&{\it $n$-dimensions}&$n=2$\\
\hline 
$T^{\alpha}$&vector&$2$&$n^2(n-1)/2$&$2$\\
$R^{\alpha\beta}$&bivector&$2$&$n^2(n-1)^{2}/4$&$1$\\
$\Sigma^{\alpha}$&vector&$n-1$&$n^2$&$4$\\
$\tau^{\alpha\beta}$&bivector&$n-1$&$n^2(n-1)/2$&$2$\\
$\eta^{\alpha}$&vector&$n-1$&$n^2$&$4$\\
\hline\hline
\end{tabular}
\end{table}

For a 2D Riemann--Cartan space, as we can take from Table 1, we have
two translation generators and one rotation generator. This allows us
to introduce a Lie (or right) duality operation, that is, a duality
with respect to the Lie--algebra indices, which maps a vector into a
covector, and vice versa:
\begin{equation}
\psi^{\star}_{\alpha} :=  \eta_{\alpha\beta}\,\psi^{\beta}\,,
\qquad \psi^{\alpha} =
\eta^{\alpha\beta}\,\psi^{\star}_{\beta} \, .   
\end{equation}
Here the completely antisymmetric tensor is defined by
$\eta_{\alpha\beta} := \sqrt{|\det
  o_{\mu\nu}|}\,\epsilon_{\alpha\beta}$, where
$\epsilon_{\alpha\beta}$ is the Levi--Civita symbol\index{Levi--Civita!symbol}
normalized to
$\epsilon_{\hat 0\hat 1}=+1$ (a circumflex on top of a number
identifies the number as an {\em an}holonomic or frame index). For
$\psi^\beta=\vartheta^\beta$ we get
$\eta_\alpha:={}\ast\!\vartheta_\alpha =\vartheta_\alpha^{\star}$,
where $\ast$ denotes the Hodge (or left) dual. Using the Lie (or
right) duality in two dimensions, we can appreciably compactify the
notation, see Table~\ref{obh:table2}.

Local Lorentz transformations are defined by the $2\times 2$
matrices $\Lambda_{\beta}{}^{\alpha}(x) \in SO(1,1)$,
\begin{equation}
  \Lambda_{\alpha}{}^{\beta}=\delta^{\beta}_{\alpha}\cosh\omega +
  \eta_{ \alpha}{}^{\beta}\sinh\omega .
\end{equation}
\begin{table}[b]\label{obh:table2}
\centering
\caption{2D geometrical objects} 
\begin{tabular}{|c|c|c|c|}
  \hline\hline {\it n=2}&{\it Valuedness}&{\it p--form}&{\it
    Components}\\ \hline $\Gamma^{\star} := (1/2)\, \eta_{\alpha\beta}
  \,\Gamma^{\alpha\beta}$ &scalar&$1$&$2$\\ 
  $t^{\alpha}:=\;{*}T^{\alpha}$&vector&$0$&$2$\\ 
  $T:=e_{\alpha}\rfloor T^{\alpha}$&scalar&$1$&$2$\\ $t^2:=
  o_{\alpha\beta}\, t^{\alpha}\, t^{\beta}$&scalar&$0$&$1$\\ 
  $R^{\star} = d\Gamma^{\star}$&scalar&$2$&$1$\\ $R:=e_{\alpha}\rfloor
  e_{\beta}\rfloor R^{\alpha\beta}$&scalar&$0$&$1$\\ \hline\hline
\end{tabular}
\end{table}

The gauge transformations of the basic gravitational field variables read
\begin{eqnarray}
  \vartheta^{\prime\alpha}&=&(\Lambda^{-1})_{\beta}{}^{\alpha}\,
  \vartheta^{\beta}
  =\vartheta^{\alpha}\cosh\omega -\eta^{\alpha}\sinh\omega,\\ 
  \Gamma^{\prime}_{\alpha}{}^\beta &=&\Lambda_{\alpha}{}^{\gamma}
  \Gamma_{\gamma}{}^{\delta}(\Lambda^{-1})_{\delta}{}^{\beta}-
  \Lambda_{\alpha}{}^{\gamma}d(\Lambda^{-1})_{\gamma}{}^{\beta} =
  \Gamma_{\alpha}{}^{\beta} + \eta_{\alpha}{}^{\beta}d\omega,\\ 
  {\rm or}\qquad\qquad
  \Gamma^{\star\prime}&=&\Gamma^{\star}-d\omega.\label{obh:Gs}
\end{eqnarray}

The curvature 2--form has only one irreducible component, and it can
be expressed in terms of the curvature scalar $R:=e_\alpha\rfloor
e_\beta\rfloor R^{\alpha\beta}$:
\begin{equation}
  R^{\alpha\beta}=- {1\over 2}\,
  R\;\vartheta^{\alpha}\wedge\vartheta^{\beta}.
\end{equation} 
In two dimensions torsion is irreducible and reduces to its vector
piece
\begin{equation}
  T^{\alpha} = - t^{\alpha}\eta,
\end{equation} 
where the vector--valued torsion zero--form  $t^\alpha$ is defined
via the Hodge dual
\begin{equation}
  t^{\alpha} := \ast T^{\alpha}.
\end{equation}

When the torsion square is not identically zero, i.e. $t^2 :=
t_{\alpha}\,t^{\alpha}\neq 0$, we call the corresponding manifold $M$
a {\it non-degenerate Riemann--Cartan spacetime}\index{Riemann--Cartan
  spacetime!non-degenerate }. In this case, using the scalar-valued
{torsion one--form} $T := e_{\alpha}\rfloor T^{\alpha}$,
we can write a coframe as 
\begin{equation}
  \vartheta^{\alpha}= - {1\over
    t^2}\left(T\,\eta^{\alpha\beta}\,t_{\beta} + \ast T\,
    t^{\alpha}\right)= - {1\over
    t^2}\left(T\,t^{\star\alpha} + \ast T\,
    t^{\alpha}\right).\label{obh:theta}
\end{equation}

Thus, the torsion one--form $T$ and its dual $\ast T$ specify a
coframe with respect to which one can expand all the 2D geometrical
objects. When $t^2\neq 0$, this coframe is non-degenerate, hence the
terminology of a non-degenerate Riemann--Cartan space. In this case,
the volume two--form can be calculated, in the non-degenerate case, as
an exterior square of the torsion one--form $T$:
\begin{equation}
  \eta := {1\over 2}\,\eta_{\alpha\beta}\,\vartheta^{\alpha}\wedge
  \vartheta^{\beta}={1\over t^2}\ast\! T\wedge T.\label{obh:vol}
\end{equation}
Defining a coframe of a 2D Riemann-Cartan spacetime in terms of the
torsion one--form turns out to be extremely convenient, and, in fact,
underlies the integrability of the 2D gravity models with and without
matter.

\section{The field equations of PGG: invariant formulation}\label{obh:sect3}

The total action of the interacting matter field $\Psi$ and the PGG
fields in two dimensions reads
\begin{equation}
W = \int \Bigl[L( \vartheta^{\alpha}, \Psi, D\Psi) +
V(\vartheta^{\alpha}, T^{\alpha}, R^{\alpha\beta})\Bigr], \label{obh:act}
\end{equation}
where the matter Lagrangian two--form $L$ will be specified later.

One can prove that torsion and curvature can enter a general
gravitational Lagrangian $V$ only in form of the scalars $t^{2}
:=o_{\alpha\beta}\,t^{\alpha}t^{\beta}$ and $R$. The gravitational
Lagrangian density is denoted by ${\cal V}:={\ast}V$.  Then the
general gauge invariant PGG Lagrangian reads
\begin{equation}
V(\vartheta^{\alpha}, T^{\alpha}, R^{\alpha\beta})=V(\vartheta^{\alpha}, 
t^2, R) = - {\cal V}(t^{2}, R)\,\eta.\label{obh:calv}
\end{equation}
The partial derivatives
\begin{equation}
  P:=-2\left({\partial{\cal V}\over \partial t^2}\right),\quad
  \kappa:=2\left({\partial{\cal V}\over \partial R}\right)\,,\label{obh:KP}
\end{equation}
i.e.\ the generalized gravitational field momenta (`excitations'),
define two functions $P=P(t^2 , R)$ and $\kappa=\kappa(t^2 , R)$ which
are assumed to be smooth and {\em nontrivial:} $P\neq 0,\,\kappa\neq
0$.

\begin{figure}[htb]
  \begin{center}
    \leavevmode
\begin{picture}(0,0)%
\epsfig{file=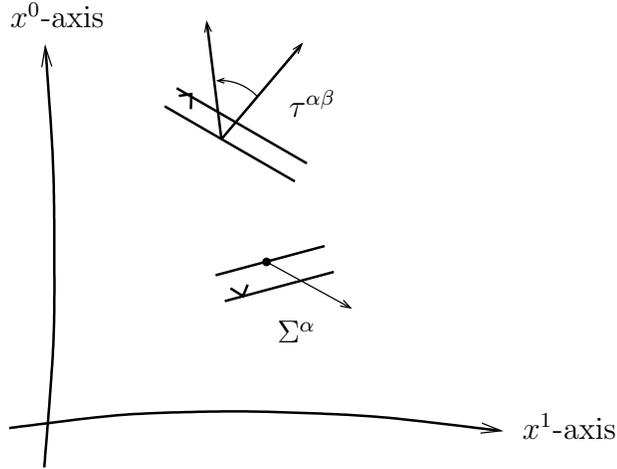}%
\end{picture}%
\setlength{\unitlength}{3947sp}%
\begingroup\makeatletter\ifx\SetFigFont\undefined%
\gdef\SetFigFont#1#2#3#4#5{%
  \reset@font\fontsize{#1}{#2pt}%
  \fontfamily{#3}\fontseries{#4}\fontshape{#5}%
  \selectfont}%
\fi\endgroup%
\begin{picture}(3247,2962)(779,-2183)
\put(811,614){\makebox(0,0)[lb]{\smash{\SetFigFont{12}{14.4}{\familydefault}{\mddefault}{\updefault}$x^0$-axis}}}
\put(2561, 54){\makebox(0,0)[lb]{\smash{\SetFigFont{12}{14.4}{\familydefault}{\mddefault}{\updefault}$\tau^{\alpha\beta}$}}}
\put(2486,-1366){\makebox(0,0)[lb]{\smash{\SetFigFont{12}{14.4}{\familydefault}{\mddefault}{\updefault}$\Sigma^\alpha$}}}
\put(4026,-1971){\makebox(0,0)[lb]{\smash{\SetFigFont{12}{14.4}{\familydefault}{\mddefault}{\updefault}$x^1$-axis}}}
\end{picture}
    \caption{
Energy-momentum current $\Sigma^\alpha$ (4 components) and spin current
$\tau^{\alpha\beta}$ (2 components) in two-dimensional spacetime.
}
  \end{center}
\end{figure}

The variational derivatives
\begin{equation}
\Sigma_{\alpha}:={\delta L \over \delta\vartheta^{\alpha}},\qquad
\tau_{\alpha\beta}:={\delta L \over \delta\Gamma^{\alpha\beta}},
\label{obh:sig}
\end{equation}
yield the energy--momentum and the spin one--forms of matter,
respectively, see Fig.3. Using the right duality, one
straightforwardly replaces the bivector--valued spin by the
scalar--valued one--form $\tau^\star:={1\over
  2}\eta_{\alpha\beta}\,\tau^{\alpha\beta}$.  Similarly, instead of
using the vector-valued energy-momentum one-form $\Sigma_{\alpha}$, it
turns out to be more convenient to introduce two scalar-valued
one-forms
\begin{equation}
  S:= t^{\alpha}\Sigma_{\alpha}, \qquad S^{\star}:=
  t_{\alpha}\,\eta^{\alpha\beta}\Sigma_{\beta}=
  t^{\star\alpha}\Sigma_{\alpha}.\label{obh:ss}
\end{equation}
Analogously to (\ref{obh:theta}), which expresses the coframe
in terms of $T$ and $\ast T$, one can rewrite the energy--momentum
current in terms of $S$ and $S^\star$:
\begin{equation}
  \Sigma_{\alpha}= {1\over t^2}\left(t_{\alpha}\,S +
    \eta_{\alpha\beta}t^\beta\,S^\star\right)={1\over
    t^2}\left(t_{\alpha}\,S +
    t^\star_\alpha\,S^\star\right).\label{obh:sigmas}
\end{equation}
If we use the Hodge star, we find by straightforward algebra
\begin{equation}\label{obh:ssrel}
  S^{\star} + \ast S
  =\ast(\vartheta^{\alpha}\wedge\Sigma_{\alpha})\ast\! T -
  \ast(\eta^{\alpha}\wedge\Sigma_{\alpha})\,T.
\end{equation}
Let us recall that $\vartheta^{\alpha}\wedge\Sigma_{\alpha}$ describes
the trace of the energy-momentum whereas
$\eta^{\alpha}\wedge\Sigma_{\alpha}$ represents its antisymmetric
part. The latter is related to the spin via the second Noether identity:
\begin{equation}
2\,d\tau^\star = \eta^{\alpha}\wedge\Sigma_{\alpha}.\label{obh:2ndNoe}
\end{equation}

The general field equations of PGG\index{PGG!general field equations}
arise from independently varying (\ref{obh:act}) with respect to the coframe
$\vartheta^{\alpha}$ and the connection $\Gamma^{\alpha\beta}$.
Remarkably, these equations can be rewritten in a completely
coordinate and gauge invariant form
\begin{eqnarray}
  d(P^{2}\,t^{2}) &=& 2P(\widetilde{\cal V}\,T+{S}),\label{obh:dpt2}\\ 
  d(P\,{\ast}T) &=& (P\,t^{2} - 2\,\widetilde{\cal V})\,\eta
  +\vartheta^{\alpha}\wedge{\Sigma}_{\alpha} ,\label{obh:dpt}\\ 
  d\kappa &=& - P\,T + 2{\tau}^\star ,\label{obh:dk}\\ 
  Pt^2(\Gamma^{\star} + du) &=& \widetilde{\cal V}\,\ast\! T + {S}^\star ,
\label{obh:gam}
\end{eqnarray}
where 
\begin{equation}
\widetilde{\cal V}:= {\cal V} + P\,t^2 - {1\over 2}\,\kappa\,R \label{obh:cv}
\end{equation}
is the so--called modified Lagrangian function and 
\begin{equation} t^2\,du:=\eta_{\alpha\beta}\,t^{\alpha}dt^{\beta}. 
\label{obh:du}\end{equation} The term $du$ in (\ref{obh:gam}) is physically 
irrelevant, since a 2D Lorentz transformation can create such an
Abelian shift, see (\ref{obh:Gs}).  In (\ref{obh:dpt2})-(\ref{obh:gam}), the
source terms of the matter field $\Psi$ are represented by $S$ and
$S^\star$, by the energy--momentum trace
$\vartheta^{\alpha}\wedge\Sigma_{\alpha}$, and by the spin
$\tau^\star$. We marked them in the field equations by letters in
boldface. Besides the gravitational field equations, we have the
matter field equation.  For matter described by a $p$--form field
$\Psi$ it reads
\begin{equation}\label{obh:matter}
{\delta L\over \delta\Psi} = {\partial L\over \partial\Psi} -
(-1)^{p}D{\partial L \over \partial D\Psi}=0.
\end{equation}

As we have seen, the system (\ref{obh:dpt2})-(\ref{obh:gam}) involves the
energy--momentum one--forms $S$ and $S^\star$ as sources.  They
satisfy the following equations which can be derived from the Noether
identities:
\begin{eqnarray}
d(PS) &=& T\wedge P(S - R\tau^\star) + 
{1\over 2}\widetilde{\cal V}\,d\tau^\star,\label{obh:ds}\\
d(PS^\star) &=& T\wedge P(S^\star + R\ast\!\tau^\star)  + 
{2\over t^2}\, S\wedge S^\star + \left(\widetilde{\cal V} - 
Pt^2\right)\vartheta^{\alpha}\wedge\Sigma_{\alpha}.\label{obh:dst}
\end{eqnarray}

In 2D, the specific feature of standard matter (scalar, spinor,
Abelian and non--Abelian gauge fields) is that the spin current is
zero:
\begin{equation}\label{obh:standard}\tau_{\alpha\beta}=0\qquad
  \hbox{\em (standard matter)}\,.\end{equation} Thus only the
canonical energy--momentum one--form $\Sigma_{\alpha}$ enters the
gravitational field equations as a source.  Moreover, in view of
(\ref{obh:2ndNoe}), it becomes symmetric:
$\eta^\alpha\wedge\Sigma_\alpha=0$. In this paper, we limit ourselves
to the discussion of massless, conformally invariant matter models.
Then we have
\begin{equation}\label{obh:tr0}
  \vartheta^{\alpha}\wedge\Sigma_{\alpha}=0\qquad\hbox{\em (massless,
    conformally invariant matter)} \,,
\end{equation} 
and the corresponding terms drop out from the field equation
(\ref{obh:dpt}) and the Noether identity (\ref{obh:dst}). Consequently, only
$S$ is left as a source. {}From now on, {\em we will specialize} to
this physically most interesting case obeying (\ref{obh:standard}) and
(\ref{obh:tr0}). Under these conditions, (\ref{obh:ssrel}) simply reduces to
\begin{equation}\label{obh:srel}
 S^{\star} + \ast S = 0.
\end{equation}

Accordingly, the complete system (\ref{obh:dpt2})-(\ref{obh:gam}) and
(\ref{obh:ds})-(\ref{obh:dst}), with (\ref{obh:standard}) and (\ref{obh:tr0}), should
be jointly solved with the matter field equation (\ref{obh:matter}).

\subsection*{Consistency check of the invariant formulation}

As it is clearly suggested by the field equation (\ref{obh:dk}), the
function $\kappa$ of the Riemann-Cartan curvature $R$ (and, in
general, of $t^2$) can be conveniently treated as one of the local
coordinates on a two--dimensional manifold $M$. However, one has
always to check the consistency of the scheme by explicitly
calculating the curvature from the connection which itself is obtained
from the field equations. This was done for the vacuum solutions of
the general PGG model in \cite{o1} and for non-vacuum solutions of
the quadratic models in \cite{o2}. Here we will demonstrate
consistency in general, for arbitrary matter sources and arbitrary
gravitational Lagrangian. We consider the non-trivial non-degenerate
case with $t^2\neq 0$.

Eq.(\ref{obh:gam}) yields the general solution for the Lorentz connection.
Starting from the definitions (\ref{obh:KP}) and using (\ref{obh:dk}), it is 
straightforward to compute the differential of the modified Lagrangian:
\begin{equation}
d\,\widetilde{\cal V} = {1\over 2}\left({1\over P}d(P^2t^2) + 
RPT\right) -R\,\tau^\star.\label{obh:dv}
\end{equation}
With the help of this relation and eqs.(\ref{obh:dpt2}), (\ref{obh:dpt}), 
(\ref{obh:dst}), (\ref{obh:ssrel}), one finds 
\begin{equation}
d[P(\widetilde{\cal V}\ast\! T + S^\star)] ={1\over P^2t^2}\,d(P^2t^2)\wedge
[P(\widetilde{\cal V}\ast\! T + S^\star)] + 
{1\over 2}R P^2T\wedge\ast T\,.\label{obh:id}
\end{equation}
The consistency proof is completed by taking the exterior differential 
of the left- and right-hand sides of equation (\ref{obh:gam}). With the
help of (\ref{obh:dv}) and (\ref{obh:id}), this yields
$$
d\Gamma^{\star}=-{1\over 2}R\,\eta .
$$

\section{Exact solutions of PGG with arbitrary gravitational 
Lagrangian}\label{obh:sect4}

The two cases of two--dimensional PGG should be treated separately,
the degenerate case with $t^2=0$ and the non--degenerate one with
$t^2\neq 0$. We will formulate our answers for matter sources obeying
(\ref{obh:standard}) and (\ref{obh:tr0}).

\subsection{Degenerate torsion solutions}\label{obh:degsol}

If $t^2 =0$, the torsion one--form is either self- or anti-self-dual,
\begin{equation}
T=\pm\ast\! T.\label{obh:duator}
\end{equation}
Then (\ref{obh:dpt}) and (\ref{obh:dk}) yield $\widetilde{\cal V}=0$. This in turn,  
with (\ref{obh:cv}) and (\ref{obh:KP}), yields
\begin{equation}
  f(R):={\cal V} - R\,{\partial{\cal V}\over\partial
    R}=0.\label{obh:constR}
\end{equation}
For a given Lagrangian ${\cal V}={\cal V}(R)$, the $t^2$-dependence
drops out because of the degeneracy, the solutions of $ f(R)=0$
determine some $R=R_1$, $R=R_2, \dots$ Therefore the curvature is
constant, $R=const$ and, by implication, also the $R$-dependent
Lorentz field momentum, $\kappa=const$. Then, from (\ref{obh:dk}), one
finds $T=0$.  Finally, an analysis of the matter field equation and of
the Noether identities shows that only trivial matter configurations
are allowed: A constant field in the case of a zero--form $\Psi$, e.g..

Summarizing, we see that the degenerate solutions of PGG
reduce to the torsionless de Sitter geometry,
\begin{equation}
  T^{\alpha}=0, \quad R= const, \quad \Psi = const,\label{obh:deg}
\end{equation}
where the constant values of the curvature are roots of equation
(\ref{obh:constR}).  Incidentally, the same turns out also to be true for
some conformally non-invariant matter, for a massive scalar field with
arbitrary self--interaction, e.g..  In the rest of the paper we will
mainly consider the non-degenerate case with $t^2\neq 0$.

\subsection{Non-degenerate vacuum solutions}

Let us now specialize to the {\it vacuum} field equations.
Accordingly, in (\ref{obh:dpt2})-(\ref{obh:dk}) we have to put $S=0$,
$\vartheta ^\alpha\wedge\Sigma_\alpha=0$, and $\tau^\star=0$. The
formal general solution is obtained as follows: Let us introduce a
coordinate system $(\kappa,\lambda)$ which is related to the torsion
1-form basis $(T,\ast T)$ via
\begin{equation}
  P\,T=-d\kappa,\qquad\qquad P\ast\! T = Bd\lambda , \label{obh:TT}
\end{equation}
with some function $B(\kappa,\lambda)$. Consequently, the volume 2--form 
is given by 
\begin{equation}
  \eta={B\over{P^2t^2}}\,d\kappa\wedge d\lambda,
\end{equation}
cf. (\ref{obh:vol}). The first equation in (\ref{obh:TT}) is simply the field
equations (\ref{obh:dk}).

Substitution of the ansatz (\ref{obh:TT}) into (\ref{obh:dpt2}) and (\ref{obh:dpt})
results in
\begin{eqnarray}
{\partial\over\partial\kappa}(P^2t^2) &=& - 2{\widetilde{\cal V}},\qquad
{\partial\over\partial\lambda}(P^2t^2) =0,\label{obh:PT}\\
{\partial\over\partial\kappa}\ln B &=&
{\partial\over\partial\kappa}\ln (P^2t^2) + {1\over P}.\label{obh:BP}
\end{eqnarray}
Formal integration of (\ref{obh:BP}) yields the solution
\begin{equation}
B=B_{0}(\lambda)\,P^2t^2\exp \left(\int {d\kappa\over P}\right)\,, 
\end{equation}
where $B_{0}(\lambda)$ is an arbitrary function of $\lambda$ only.

Provided the gravitational Lagrangian $V$, and hence $P$, is smooth,
there always exists a solution of the first order ordinary
differential equations (\ref{obh:PT}). This describes
$P^2t^2$ as a function of $\kappa$ and $\lambda$, thus completing our
formal demonstration of the integrability of the general
two-dimensional vacuum PGG. The complete non-degenerate vacuum
solution is evidently of the black hole type with the
metric\index{PGG!vacuum solution of}\index{PGG!black hole type solution of}
\begin{equation}
g=-{d\kappa^2 \over P^2t^2} + 
P^2t^2 \exp\left(2\int {d\kappa\over P}\right)d\lambda^2.\label{obh:vac-met}
\end{equation}
Here, without restricting generality, we put $B_0=1$.  {\em Torsion}
and curvature for our solution are obtained by inverting the relations
$P=P(t^2,R),\kappa=\kappa(t^2,R)\rightarrow t^2=t^2(P,\kappa),
R=R(P,\kappa)$. For the solution to be unique, one must assume the
relevant Hessian (${\partial^2{\cal V} \over \partial t^2 \partial
  t^2},{\partial^2{\cal V}\over \partial R\partial R}$) to be
non-degenerate. It is straightforward to derive from (\ref{obh:gam}) the
{\em curvature} scalar of the general solution:
\begin{equation}
R={P^2t^2 \over B}{\partial \over \partial\kappa}\left({B\over P^2t^2}
{\partial\over \partial\kappa}(P^2t^2) \right). 
\end{equation}

The position of the horizon(s) is evidently determined by the zeros of
the metric coefficient $g_{\lambda\lambda}=P^2t^2\exp\left( 2\int
  d\kappa/P\right)$. It is impossible to say more without explicitly
specializing the gravitational Lagrangian. An important particular
case is represented by quadratic PGG which will be discussed in
the next two sections.

\section{Exact vacuum solutions of PGG with quadratic gravitational 
Lagrangian}\index{PGG!vacuum solutions for quadratic Lagrangian}
\label{obh:sect5}

Let us now analyze two-dimensional PGG with a gravitational Lagrangian
{\it qua\-drat\-ic} in torsion and curvature,\index{PGG!quadratic
  gravitational Lagrangian}
\begin{equation}
V = - \left({a\over 2}\,T_\alpha {}*\!T^\alpha 
+{1\over 2}\, R^{\alpha\beta}\eta _{\alpha\beta}+{b\over 2}\,
R_{\alpha\beta }{}*\!R^{\alpha\beta }\right) - \Lambda\,\eta .\label{obh:Lgr}
\end{equation}
Here $a$, $b$, and $\Lambda$ are the coupling constants. Using (\ref{obh:Lgr})
in (\ref{obh:KP}), we find:
\begin{equation}
P=a,\qquad\kappa =b\,R -1.\label{obh:KP2}
\end{equation}
The modified Lagrangian function reads
\begin{equation}
\widetilde{\cal V}={a\over 2}\,t^2 - {b\over 4}\,R^2 + \Lambda .\label{obh:wv}
\end{equation}

We will concentrate here on {\it vacuum solutions}. The general scheme
for an arbitrary Lagrangian $V$ was given in the previous section. The
degenerate solutions (\ref{obh:deg}) are de Sitter spacetimes:
\begin{equation}
T^\alpha =0,\qquad R=\pm R_{\rm dS},\qquad
R_{\rm dS}:= 2\sqrt{\Lambda\over b}.\label{obh:desit}
\end{equation}
Also the non-degenerate solutions can be easily obtained. Substituting
(\ref{obh:KP2})-(\ref{obh:wv}) into (\ref{obh:PT}), we explicitly find for the
scalar {\em torsion} square
\begin{equation}
-t^2=2M_0\, e^{-bR/a} - {b\over 2a}\,R^2 + R + {2\Lambda\over a} - 
{a\over b}.\label{obh:t2}
\end{equation}

The integration constant $M_0$\index{PGG!mass of non-degenerate vacuum
  solution} has the physical meaning of the mass of a point-like
source. This can be derived from the existence of the timelike Killing
vector field $\zeta=\partial_\lambda$ which yields the conserved
energy--momentum 1--form
\begin{equation}
\varepsilon_{\rm RC}:=\zeta^\alpha\Sigma_\alpha = -e^{bR/a}\left({1\over 
2}\,d(t^2) + {1\over a^2}{\widetilde{\cal V}}\,d\kappa\right).\label{obh:deps}
\end{equation}
This 1--form is {\it strongly} conserved, i.e.\ $d\varepsilon_{\rm
  RC}=0$ even when the field equations are not fulfilled. One can
verify that
\begin{equation}
\varepsilon_{\rm RC} = dM,\qquad
M:=-{e^{bR/a}\over 2}\left(t^2 - {b\over 2a}R^2 + R + {2\Lambda\over a}
- {a\over b}\right).\label{obh:M}
\end{equation}
When the {\it vacuum} field equations are satisfied, $\varepsilon_{\rm
  RC} =0$, and thus $M=M_0$.

\subsection*{Black hole structure of non-degenerate spacetimes}\label{obh:bhgeom}

The degenerate solutions are torsionless de Sitter spacetimes with
constant Riemannian curvature $R_{\rm dS}$, see (\ref{obh:desit}).
The properties of the non-dege\-nerate solutions are obviously defined
by the values of the coupling constants $(a, b, \Lambda)$ and by the
value of the first integral $M_0$. Let us denote
\begin{equation}
M_{\pm}:={e^{\mp{b}R_{\rm dS}/a}\over 2}\left({a\over b} \pm R_{\rm dS}
\right).\label{obh:Mpm}
\end{equation}
We recognize that always $M_{-}\leq M_{+}$, equality is achieved only
for vanishing cosmological constant: $M_{-}= M_{+}=a/(2b)$ for $R_{\rm dS}=
\Lambda=0$. For sufficiently large $\Lambda$, namely when $\Lambda >
a^2/(4b)$ (or, equivalently, $R_{\rm dS} > a/b$) one finds negative $M_{-}$,
otherwise $M_{\pm}\geq 0$. The special case $\Lambda=a^2/(4b)$ in a de 
Sitter gauge gravity model was discussed in \cite{sol1,sol3}
(then $R_{\rm dS}=a/b$ and $M_{-}=0$).

As we already know, the spacetime metric is given by the line element
(\ref{obh:vac-met}) with (\ref{obh:KP2}) and (\ref{obh:t2}) inserted. Clearly, instead 
of $\kappa$, we can use the scalar curvature $R$ as a spatial coordinate.
Another convenient choice is a ``radial" coordinate 
\begin{equation}
r:=\exp\left({b\over a}R\right).\label{obh:r-coord}
\end{equation}

The meaning of the quantities (\ref{obh:Mpm}) becomes clear when we analyze
the metric coefficient
\begin{equation}
g_{\lambda\lambda}=-2M_0\, e^{bR/a} + e^{2bR/a}\left({b\over 2a}R^2 - R 
- {2\Lambda\over a} + {a\over b}\right).\label{obh:g00}
\end{equation}
Its zeros define the positions of the horizons\index{PGG!horizons of
  exact solutions} $R_h$,
\begin{equation}
g_{\lambda\lambda}(R_h)=0.\label{obh:hor}
\end{equation}
At $R=-\infty$, we have $g_{\lambda\lambda}=0$. However, this point is
not a horizon but a true singularity with infinite curvature. This
corresponds to $r=0$ which one can consider as the position of a central
point-like source mass. Such a singularity is, in general,
hidden by the horizons.

\begin{figure}
\epsfxsize=\hsize \epsfbox{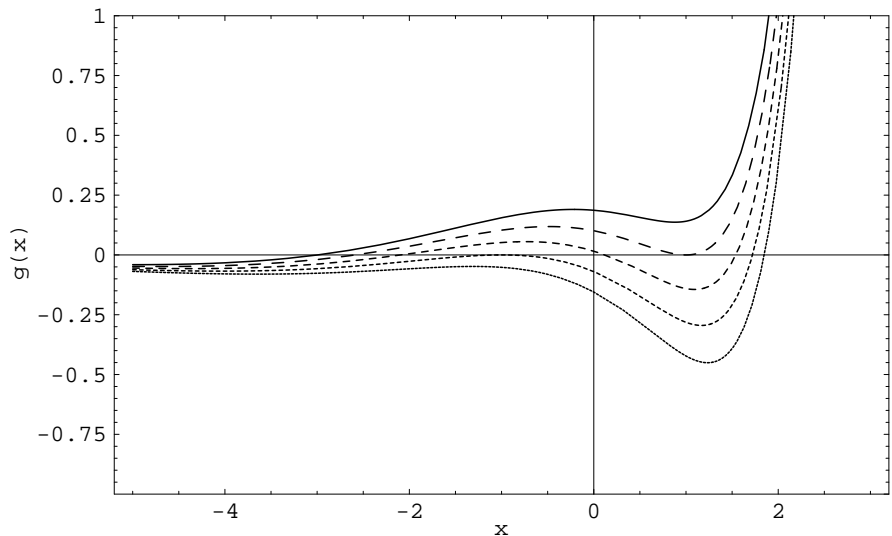}
\caption[Metric coefficient $g(x):=g_{\lambda\lambda}/R_{dS}$ as a function
of $x:=R/R_{dS}$: Moving from upper to lower curves corresponds to the cases
(i)-(v), respectively.]
{\label{obh:fig4}Metric coefficient $g(x):=g_{\lambda\lambda}/R_{dS}$ as a function
of $x:=R/R_{dS}$: Moving from upper to lower curves corresponds to the cases
(i)-(v), respectively.}
\end{figure}


The position, the number, and the type of horizons are completely
determined by the total mass $M_0$. We can distinguish five
qualitatively different configurations:
\begin{itemize}
\item[(i)] $\quad M_0<M_{-}$: one horizon at
\begin{equation}
R_h < -R_{\rm dS}.
\end{equation}
\item[(ii)] $\quad M_0=M_{-}$: two horizons at 
\begin{equation}
R_{h_1} < -R_{\rm dS},\qquad R_{h_2}=R_{\rm dS}.
\end{equation}
\item[(iii)] $\quad M_{-}<M_0<M_{+}$: three regular horizons at 
\begin{equation}
R_{h_1} < -R_{\rm dS} < R_{h_2} < R_{\rm dS},\qquad R_{h_{c}}>R_{\rm dS}.
\end{equation}
\item[(iv)] $\quad M_0=M_{+}$: two horizons at
\begin{equation}
R_{h_1} = -R_{\rm dS},\qquad R_{h_2}>R_{\rm dS}.
\end{equation}
\item[(v)] $\quad M_0>M_{+}$: one regular horizon at
\begin{equation}
R_h > R_{\rm dS}.
\end{equation}
\end{itemize}
Here we have assumed $M_0 > 0$ which is physically reasonable. For
zero or negative values of $M_0$, the smallest horizons disappear for
the cases (i), (ii), and (iii). In Fig.~\ref{obh:fig4}, the function 
(\ref{obh:g00}) is depicted for $\Lambda={a^2\over 16b}$ and positive
values of $M_0$.

The geometry of these solutions is obtained by an appropriate gluing
of a charged black hole to a de Sitter spacetime.  In (iii), the
largest $R$, namely $R_{h_c}$, describes the cosmological event
horizon which hides de Sitter singularity at $R=\infty$ ($r=\infty$)
from an observer at $R < R_{h_c}$. Other horizons, with $R_{h_{1,2}}$,
describe the charged black hole. The cases (ii) and (iv) correspond to
the extremal Reissner-Nordstr\"om black hole.

The Hawking temperature\index{PGG!Hawking temperature of black hole
  solution} of the black holes is related to their surface gravity.
The latter can be straightforwardly calculated from the knowledge of
the timelike Killing vector $\zeta=\partial_\lambda$:
\begin{equation}
\sqrt{-{1\over 2}(\nabla_i\zeta_j)(\nabla^i\zeta^j)}
\;\vline\,{\hbox{\raisebox{-2.5ex}{\scriptsize $R=R_h$}}}={b\over 2a^2}
\left(R_h{}^2 - R_{dS}{}^2\right).
\end{equation}
The temperature vanishes for the extremal black hole configurations
(ii) and (iv).

\begin{figure}
\epsfxsize=\hsize \epsfbox{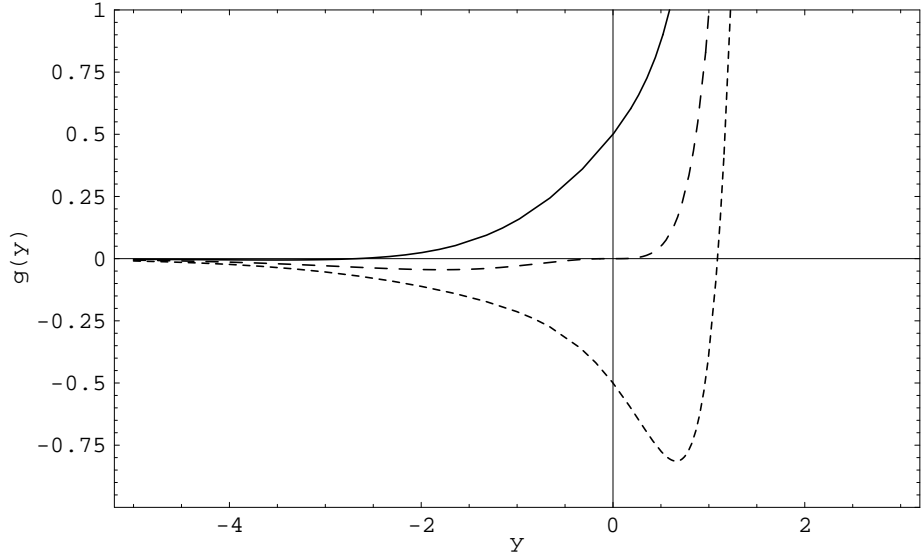}
\caption[Case $R_{dS}=\Lambda =0$: Metric coefficient $g(y):=
g_{\lambda\lambda}b/a$ as a function of $y:=Rb/a$. Upper, central, and
lower curves describe the cases (i), (ii)-(iv), and (v),
respectively.]  {\label{obh:fig5}The case $R_{dS}=\Lambda =0$: The metric
  coefficient $g(y):= g_{\lambda\lambda}b/a$ is plotted as a function of
  $y:=Rb/a$. Upper, central, and lower curves describe the cases (i),
  (ii)-(iv), and (v), respectively.}
\end{figure}

It seems worthwhile to mention that $g_{\lambda\lambda}$ reaches a
local maximum at $R_{h_1}$ [case (iv)], and a local minimum at
$R_{h_2}$ [case (ii)], when $M_{-}\neq M_{+}$. For $M_{-}=M_{+}$, the
three cases (ii), (iii), and (iv) degenerate to a configuration with a
horizon at $R_{h_1}= R_{h_2}=0$, which is an inflexion point of
$g_{\lambda\lambda}$. The qualitative behavior is given in
Fig.~\ref{obh:fig5} for positive values of $M_0$.

\section{Quadratic PGG with massless matter}\label{obh:sect6}

\subsection{Massless fermions}

Dirac spinors\index{PGG!massless Dirac spinors in} in two dimensions
have two (complex) components,
\begin{equation}
\psi= \pmatrix{ \psi_1 \cr \psi_2 }.\label{obh:dir}
\end{equation}
The spinor space at each point of the spacetime manifold is related to
the tangent space at this point via the spin-tensor objects.  In the
approach with Clifford--algebra valued forms, the central object is
the Dirac one--form
\begin{equation}
\gamma=\gamma_{\alpha}\vartheta^{\alpha}\, ,\label{obh:cliff}
\end{equation}
which satisfies
\begin{equation}
\gamma\otimes\gamma=g,\ \ \ \ \
\gamma\wedge\gamma=-2\gamma_5 \eta.\label{obh:gg}
\end{equation}
The $\gamma_5$ matrix is implicitly defined by
\begin{equation}
  \ast\gamma =\gamma_5 \gamma.\label{obh:g5}
\end{equation}
We are using the following explicit realization of the Dirac one--form:
\begin{equation}
\gamma := \pmatrix{0 & -(\vartheta^{\hat{0}}-\vartheta^{\hat{1}}) \cr
(\vartheta^{\hat{0}}+\vartheta^{\hat{1}}) & 0 \cr}
.\label{obh:mat}
\end{equation}
The Dirac matrices $\gamma^\alpha$ satisfy the usual identity $\gamma^{\alpha}
\gamma^{\beta}+\gamma^{\beta}\gamma^{\alpha}=2g^{\alpha\beta}$.

The gauge and coordinate invariant Lagrangian two--form for the massless 
Dirac spinor field can be written in the form
\begin{equation}
L={i\over 2}(\bar{\psi}\,\gamma\wedge d\psi + d\bar{\psi}\wedge\gamma\,\psi).
\label{obh:ld}
\end{equation}
It is well known that in two dimensions there is no interaction of spinors 
with the Lorentz connection, and the above Lagrangian contains ordinary 
exterior differentials and not the covariant ones. Nevertheless, the theory 
is explicitly invariant under local Lorentz rotations. 

The (Dirac) field equation is obtained from the variation of $L$ with 
respect to $\bar{\psi}$ and reads
\begin{equation}
\gamma\wedge d\psi - {1\over 2}(d\gamma)\psi = 0.\label{obh:deq}
\end{equation}
The degenerate case was described in Sec.\ \ref{obh:degsol}. Here we
assume that $t^2\neq 0$. Thus the one--forms $T$ and $\ast T$ can be
treated as the coframe basis in a two--dimensional Riemann-Cartan
spacetime. In the explicitly gauge invariant approach, it is
convenient to define, instead of the original spinor (\ref{obh:dir}), two
complex (four real) Lorentz invariant functions:
\begin{equation}
\varphi_{1}=u_1+iu_2:=\sqrt{(t^{\hat{0}} + t^{\hat{1}})}\,\psi_{1},\quad
\varphi_{2}=v_1+iv_2:=\sqrt{(t^{\hat{0}} - t^{\hat{1}})}\,\psi_{2}.\label{obh:ind}
\end{equation}
The Dirac equation (\ref{obh:deq}) yields for the real variables 
$u_A, v_A,\ A=1,2$ (using (\ref{obh:theta}))
\begin{equation}
d\left[{u_Au_B\over t^2}(T-\ast T)\right] =0,\quad\quad
d\left[{v_Av_B\over t^2}(T+\ast T)\right] =0.\label{obh:df}
\end{equation}
This can be immediately integrated. In particular, the Poincar\'e
lemma (locally) guarantees the existence of two real functions. We
denote them by $x$ and $y$, such that
\begin{equation}
{|\varphi_{1}|^2\over t^2}(T-\ast T) = dx,\quad\quad
{|\varphi_{2}|^2\over t^2}(T+\ast T) = dy.\label{obh:dx}
\end{equation}
This evidently provides local null coordinates $(x,y)$ for the
spacetime manifold. Introducing the phases of the complex spinor
components explicitly,
\begin{equation}
\varphi_{1}=|\varphi_{1}|e^{i\alpha}, \quad
\varphi_{2}=|\varphi_{2}|e^{i\beta}, \label{obh:phase}
\end{equation}
we find, using (\ref{obh:df}), that these phases depend only on one
of the above variables,
\begin{equation}
\alpha =\alpha(x),\quad \beta =\beta(y).\label{obh:solab}
\end{equation}
This construction gives the general exact solution of the massless
Dirac equation in an arbitrary two--dimensional Riemann--Cartan spacetime.

The energy--momentum one--form is straightforwardly obtained,
\begin{equation}
S=A_{1}(T-\ast T) + A_{2}(T+\ast T),\label{obh:sgen}
\end{equation}
where we denoted
\begin{equation}
A_{1}:=-\,{d\alpha \over dx}{|\varphi_{1}|^4 \over t^2}, \quad
A_{2}:=-\,{d\beta \over dy}{|\varphi_{2}|^4 \over t^2}.\label{obh:a12}
\end{equation}
In (\ref{obh:sgen}), we have $S^{\star}=-\ast S$, well in accordance with
(\ref{obh:srel}).

\subsection{Massless bosons}

Let us now turn to a gravitationally coupled massless scalar
field\index{PGG!massless scalar (Klein-Gordon) fields in} $\phi$ with
the Lagrangian two--form
\begin{equation}
L= - {1\over 2}\,d\phi\wedge{}\ast d\phi .\label{obh:Lphi}
\end{equation}
Variation with respect to $\phi$ yields the Klein-Gordon equation:
\begin{equation}
\ast d\ast\! d\phi = 0. \label{obh:KG}
\end{equation}

In non-degenerate spacetimes with $t^2\neq 0$, we can 
use the torsion coframe basis and can write, in the most general case,
\begin{equation}
d\phi = \Phi_{1}(T-\ast T) + \Phi_{2}(T + \ast T),\label{obh:dp}
\end{equation}
with some functions $\Phi_{1,2}$. We substitute (\ref{obh:dp}) into the
Klein-Gordon equation (\ref{obh:KG}). Then it turns out that locally there
exists such a scalar function $z$ that
\begin{equation}
\Phi_{1}(T-\ast T) - \Phi_{2}(T + \ast T) = d\,z.\label{obh:genKG}
\end{equation}
This describes a general solution of the Klein--Gordon equation. 

For the energy-momentum one--form $S$ we get, similarly to
(\ref{obh:sgen}),
\begin{equation}
S=A_{1}(T-\ast T) + A_{2}(T + \ast T),\label{obh:sc}
\end{equation}
where now
\begin{equation}
A_{1}= - t^{2}\Phi_{1}^2,\quad A_{2}= - t^{2}\Phi_{2}^2. \label{obh:as12}
\end{equation}

\subsection{Chiral solutions}

Both, the massless Dirac equation and the massless Klein--Gordon
equation, admit chiral solutions. For fermions chirality means that
only one component of the spinor field is nontrivial. For bosons
chirality can be formulated in terms of self- or anti-self-duality of
the ``velocity'' one--form $d\phi$. In both cases, the field equations
describe right- or left-moving configurations.  In this section we
describe the corresponding gravitational field for the quadratic PGG
model (\ref{obh:Lgr}).

Let us assume that $\varphi_{2}=\psi_{2}=0$ for the fermion and
$\Phi_{2}=0$ for the boson field. Then $A_{2}=0$ in (\ref{obh:a12}) as
well as in (\ref{obh:as12}). Hence the energy--momentum one--form $S$ is
anti-self-dual $S^{\star}=S$. These are chiral configurations.

The integrals (\ref{obh:dx}) and (\ref{obh:genKG}),
\begin{equation}
T-\ast T = {t^2\over |\varphi_{1}|^2}\,d\,x\quad\ {\rm (spinor)},\quad
T-\ast T = {1\over\Phi_1}\,d\,z \quad\ {\rm (scalar)},\label{obh:ttchi}
\end{equation}
together with the equations (\ref{obh:dk}), (\ref{obh:KP2}), suggest a natural
interpretation of the variables $R$ and $x$ (or $R$ and $z$,
respectively,) as two local spacetime coordinates. Clearly, $x$ and
$z$ are different in each case, but we can unify the two problems
without risk of confusion. For the torsion coframe the equations
(\ref{obh:ttchi}) and (\ref{obh:dk}) explicitly yield 
\begin{equation}
\ast T= -\left({b\over a}dR + Bdx\right),\qquad T=-{b\over a}dR,\label{obh:cof1}
\end{equation}
for the vacuum case, see (\ref{obh:TT}). Hence the volume two--form reads
\begin{equation}
\eta = {1\over t^2}\ast\! T\wedge T = {bB\over at^2}\,dx\wedge dR,\label{obh:vol1}
\end{equation}
and the spacetime metric is given by
\begin{equation}
g={1\over t^2}\left[\left(Bdx + {b\over a}dR\right)^2 - 
{b^2\over a^2}dR^2\right].\label{obh:int1}
\end{equation}
Here we use the unifying notation
\begin{equation}
B= {t^2\over |\varphi_{1}|^2}.\label{obh:bb}
\end{equation}
for fermions, while for bosons this function relates the two
coordinate systems via $dz=\Phi_{1}B\,dx$. 

The spacetime geometry is completely described when one solves the
field equations (\ref{obh:dpt2})-(\ref{obh:dk}), (\ref{obh:ds})-(\ref{obh:dst}), thus
finding the functions $t^2$ and $B$ explicitly. By means of
(\ref{obh:sgen}) and (\ref{obh:sc}), the energy--momentum one--form turns out
to be
\begin{equation}
S=A\,dx,\label{obh:aa}
\end{equation}
where 
\begin{equation}
A:= - |\varphi_{1}|^{2}{d\alpha\over dx}\quad\ {\rm (spinor)},\quad
A:= - t^{2}B\Phi_{1}^2 \quad\ {\rm (scalar)}.\label{obh:aaa}
\end{equation}
Substituting (\ref{obh:aa}), (\ref{obh:cof1}) into (\ref{obh:dpt2})-(\ref{obh:dk}) and
(\ref{obh:ds}), one finds
\begin{eqnarray}
{\partial t^2 \over \partial R}=-\,{2b\over a^2}\widetilde{\cal V},
\qquad {\partial t^2 \over \partial x}&=& {2\over a}A,\label{obh:dtr}\\
{1\over b}{\partial \ln B \over \partial R} + {1\over a} - 
{2\over a^2 t^2}\widetilde{\cal V} &=& 0,\label{obh:dbr}\\
{\partial A \over \partial R} + {b\over a}A &=&0.\label{obh:dar}
\end{eqnarray}
The equation (\ref{obh:dst}) is redundant. This can be compared with the
vacuum case (\ref{obh:PT}),(\ref{obh:BP}).

The system (\ref{obh:dtr})-(\ref{obh:dar}) is solved by
\begin{eqnarray}
A&=&f(x)\,e^{-bR/a},\label{obh:A}\\
B&=&B_{0}(x)\,t^2e^{bR/a},\label{obh:B}
\end{eqnarray}
\begin{equation}
-t^2 = 2\,M(x)\,e^{-bR/a} - {b\over 2a}\,R^2 + R 
+ {2\Lambda \over a} - {a\over b},\label{obh:tsq}
\end{equation}
where
\begin{equation}
M(x):=-\,{1\over a}\int f(x)dx,\label{obh:Mx}
\end{equation}
with the arbitrary functions $f(x)$ and $B_{0}(x)$. Without loss of
generality we can put $B_{0}=1$ since a redefinition of $x$ is always
possible. For completeness, let us write down the Lorentz connection.
Inserting (\ref{obh:A})-(\ref{obh:tsq}) into (\ref{obh:gam}), we find
\begin{equation}
\Gamma^{\star}=d\widetilde{u} + {e^{bR/a}\over 2}\left(R 
- {a\over b}\right)dx,\label{obh:gamsol1}
\end{equation}
where $\widetilde{u}$ is a pure gauge contribution. 

The gravitational field defined by (\ref{obh:A})-(\ref{obh:tsq}) has the same
form for chiral fermionic and bosonic sources. However, the function
$f(x)$ is different for each particular physical source.

For {\it fermions} combining (\ref{obh:B}), (\ref{obh:bb}), (\ref{obh:A}), and (\ref{obh:aaa}), 
we find
\begin{equation}
|\varphi_{1}|^2 = e^{-bR/a},
\qquad  f(x)=-{d\alpha \over dx}.\label{obh:fx}
\end{equation}
Hence the solution for the chiral fermion field, in terms of its
invariant complex component, reads
\begin{equation}
\varphi_{1}=\exp\left(-{b\over 2a}R + i\alpha(x)\right), \label{obh:chsol}
\end{equation}
whereas the metric is described by (\ref{obh:int1}) with $B$ as specified
in (\ref{obh:B}) and
\begin{equation}
M(x)={\alpha(x)\over a} + M_0,\label{obh:tsq1}
\end{equation}
where $M_{0}$ is an arbitrary integration constant. 

For {\it bosons}, combining (\ref{obh:A})-(\ref{obh:B}) with (\ref{obh:aaa}) and 
(\ref{obh:dp}), one finds
\begin{equation}
f(x)= - (\Phi_{1}B)^2 = -\left({d\phi \over dx}\right)^2,\label{obh:fsc}
\end{equation}
and the scalar field $\phi(x)$ remains an arbitrary function of $x$.

The physical meaning of the solutions obtained is clear. In Sec.\ 
\ref{obh:sect5} it was shown that the structure of a static black hole in
vacuum is determined by the value of the total mass $M_0$ entering the
torsion square (\ref{obh:t2}). The massless chiral (fermionic and bosonic)
matter contributes a variable ``mass" $M(x)$ to the torsion square
(\ref{obh:tsq}). As a result, the black hole in general becomes non-static
(\ref{obh:int1}). 

One can illustrate this process of a restructuring of a black hole by
matter falling into it \cite{cal,sol6}.  Let us consider
$f(x)=-m\,\delta\left({x-x_0\over a}\right)$. In view of (\ref{obh:aa})
and (\ref{obh:A}), this function describes a point-like ``impulse" of
matter: the field energy is zero everywhere except for a single moving
point (recall that $x$ is a null coordinate).  Then, for (\ref{obh:Mx}),
one obtains
\begin{equation}
M(x)= M_0 + m\,\theta(x-x_0).
\end{equation}
In the region $x<x_0$ we have a static black hole with mass $M_0$, 
whereas for $x>x_0$ its mass increases to $M_0 + m$. 

\subsection{Conformally non-invariant matter}

The above results are restricted to the chiral case and the
conformally invariant massless matter sources. Some remarks are
necessary for the more general cases. The non--chiral solutions were
obtained in \cite{o2} for fermionic and in \cite{o4} for bosonic
matter. In general, the resulting system cannot be integrated
analytically, and a numeric analysis is needed.  It is possible,
though, to obtain exact analytic solutions for certain models with a
complicated matter content: a nonlinear spinor field interacting with
scalars, e.g., see the discussion of the instanton type solutions in
\cite{akd1}-\cite{akd3}.

The lack of conformal invariance does not always lead to serious
difficulties. Let us consider Yang-Mills theory, for example, with an
arbitrary in general non-Abelian gauge group. The dynamical variable
is the gauge potential or, equivalently, the Lie algebra-valued
connection one--form $A^B$. The Yang-Mills Lagrangian is constructed
from the corresponding gauge field strength two--form $F^{B}$:
\begin{equation}
L_{YM}= -\,{1\over 2}\, F^{B}\wedge\ast F_{B}.\label{obh:LYM}
\end{equation}
The energy--momentum current attached to (\ref{obh:LYM}) reads 
\begin{equation} 
\Sigma_{\alpha}: = e_{\alpha}\rfloor L_{YM} +  
(e_{\alpha}\rfloor F^B)\wedge\ast F_B 
=-{1\over 2}f^2 \,\eta_{\alpha}\,, 
\end{equation}
where $f^2:=f^{B}f_{B},\ f_{B}=\ast F_{B}$. As in the previous cases,
the spin current vanishes, $\tau_{\alpha\beta}=0$, since the Lorentz
connection does not couple to the Yang--Mills potential.

Observe that the energy--momentum trace, in contrast to four 
dimensions, does {\it not} vanish:
\begin{equation}
\vartheta^{\alpha}\wedge\Sigma_{\alpha} =-f^2\eta\neq 0.\label{obh:trace}
\end{equation}
The results described in Sec.\ \ref{obh:sect3} refer only to the
conformally invariant case. Thus one needs a proper generalization for
the case (\ref{obh:trace}). Quite fortunately, the situation is greatly
simplified due to the constancy of $f^2$. Although, of course, the Lie
algebra--valued scalar field $f_{A}$ is not constant in view of the
nonlinear nature of the Yang-Mills equations
\begin{equation}
D\,\ast F_A =d\,f_A + c_{ABC} A^B\,f^C =0,\label{obh:YME}
\end{equation}
obviously its square is conserved: $f^2=const$. 

As a result, the gravitational field equations (\ref{obh:dpt2})-(\ref{obh:dk})
are modified by a following simple shift of the cosmological constant:
\begin{equation}\label{obh:cosm}
\Lambda \rightarrow\widetilde\Lambda= \Lambda + {1\over 2}\,f^2. 
\end{equation}
In particular, the complete integrability of the vacuum system is not
disturbed by Yang--Mills matter, and one again recovers the static
black hole solutions described in Sec.\ \ref{obh:sect5}.

\section{Black hole solution for general dilaton gravity}\label{obh:sect7}

In this section we demonstrate that our method also successfully works
in the string motivated dilaton models, cf.
\cite{cal,banks,nappi1,nappi2,nappi3,str2,str3,russo,russo2}.

Let us denote the purely Riemannian curvature scalar by a tilde:
$\widetilde{R}$. In general, the same notation is used for all
geometrical objects and operations which are defined by the
torsion-free Riemannian connection $\widetilde{\Gamma}^{\alpha\beta}$
(Christoffel symbols).  Let be given, in two dimensions, the
gravitational potential $\vartheta^\alpha$ on one side and a
scalar field $\Phi$ and a Yang--Mills potential $A^B=A^B_i dx^i$ on
the other, the matter side. These fields are interacting with each
other. A corresponding general Lagrangian two--form reads
\begin{equation}
V_{\rm dil}=\eta\left({\cal F}(\Phi)\widetilde{R} + {\cal G}(\Phi) 
(\partial_\alpha\Phi)^2 + {\cal U}(\Phi) + {\cal J}(\Phi)
(F^B_{ij})^2\right).\label{obh:Vdil}
\end{equation}
Here the kinetic terms are constructed from
$\partial_\alpha\Phi:=e_\alpha\rfloor d\Phi$ and $F^B={1\over
  2}F^B_{ij}dx^i\wedge dx^j$, respectively.  For the string motivated
dilaton models, the coefficient functions read:\index{PGG!Lagrangian
  for dilaton gravity}
\begin{equation}
{\cal F}(\Phi)=e^{-2\Phi},\quad
{\cal G}(\Phi)=\gamma\, e^{-2\Phi},\quad
{\cal U}(\Phi)=e^{-2\Phi}U(\Phi),\quad
{\cal J}(\Phi)={-e^{(\epsilon - 2)\Phi}\over 4},\label{obh:strfun}
\end{equation}
where $\gamma=4$ and $U(\Phi)=c$ in the tree approximation of string
theory.  A number of physically interesting models correspond to
different values of $\gamma, \epsilon$, and $U(\Phi)$.

\subsection{Main result}\label{obh:sol}

Locally one can always treat the scalar function $\Phi$ as a
coordinate on a two-dimensional spacetime manifold. Denote the second
coordinate by $\lambda$.

In terms of the local coordinates $(\lambda, \Phi)$, the metric of the
general solution of the gravitational field equation of the model
(\ref{obh:Vdil}) reads
\begin{equation}
ds^2 = - 4\,h(\Phi)\,e^{-2\nu(\Phi)}\,d\lambda^2 
+ {({\cal F}')^2\over h(\Phi)}d\Phi^2,\label{obh:dsdil}
\end{equation}
where 
\begin{eqnarray}
h(\Phi)&=&e^{\nu(\Phi)}\,\left({M_0\over 2} + \int\limits_{}^{\Phi}
d\varphi\left({\cal U}(\varphi) + {Q_0^2\over 2{\cal J}(\varphi)}\right)
{\cal F}'(\varphi)\,e^{-\nu(\varphi)}\right),\label{obh:dilh}\\
\nu(\Phi)&=&\int\limits_{}^{\Phi}d\varphi{{\cal G}(\varphi)\over
{\cal F}'(\varphi)}.\label{obh:dilnu}
\end{eqnarray}
Here $M_0$ and $Q_0^2$ are the integration constants which are related
to the total mass and the (squared) charge of a solution.
For completeness, let us give the solution for the Yang--Mills field:
\begin{equation}
F^B= f^B\,\eta,\qquad f^Bf_B=\left({Q_0\over {\cal J}(\Phi)}\right)^2.
\end{equation}

The proof of this result, see below, is obtained by the method
developed for two-dimensional PGG.

\subsection{Dilaton models and PGG}

In two dimensions, torsion is represented by its trace one-form
$T:=e_\alpha\rfloor T^\alpha$, see Table 2. Then the Riemann-Cartan
connection decomposes into the Riemannian and post-Riemannian pieces
as follows:
\begin{equation}
  \Gamma_{\alpha\beta} =\widetilde{\Gamma}_{\alpha\beta} -
  \vartheta_\alpha\,e_\beta\rfloor T +
  \vartheta_\beta\,e_\alpha\rfloor T
  =\widetilde{\Gamma}_{\alpha\beta}-\eta_{\alpha\beta}\ast\!
  T.\label{obh:decgam}
\end{equation}
By differentiation we find a corresponding decomposition of the
Riemann-Cartan curvature. For the Hilbert-Einstein two-form this yields
\begin{equation}
  R^{\alpha\beta}\eta_{\alpha\beta} = \widetilde{R}^{\alpha\beta}
  \eta_{\alpha\beta} + 2\,d\ast\! T.\label{obh:HE}
\end{equation}

Let us consider the ``scalar-tensor" type PGG model with
the Lagrangian 
\begin{equation}
V_0=-{1\over 2}\left[\xi(\Phi)\,T_\alpha\wedge\ast T^\alpha +
\omega(\Phi)\, R^{\alpha\beta}\eta_{\alpha\beta}\right].\label{obh:V0}
\end{equation}
Here $\xi(\Phi)$ and $\omega(\Phi)$ are the scalar functions which
describe a variable gravitational ``constant" \`a la Jordan and
Brans-Dicke. Variation of (\ref{obh:V0}) with respect to the connection
yields the field equation
\begin{equation}
\xi(\Phi)\,T = \omega'(\Phi)\,d\,\Phi.\label{obh:TdP}
\end{equation}
Hereafter the prime denotes derivative with respect to $\Phi$. 

As we see, the torsion trace turns out to be an exact one-form, and the
scalar dilaton field plays a role of its generalized potential. Substituting
(\ref{obh:TdP}) into (\ref{obh:HE}) and (\ref{obh:V0}), one finds (dropping an
exact form):
\begin{eqnarray}
V_0&=&-{1\over 2}\left(\omega\widetilde{R}^{\alpha\beta}\eta_{\alpha\beta}
- {(\omega')^2\over \xi}\,d\Phi\wedge\ast d\Phi\right)\\
&=&\eta\left({\omega\over 2}\widetilde{R} + {(\omega')^2\over 2\xi}
(\partial_\alpha\Phi)^2\right).\label{obh:eqV0}
\end{eqnarray}
This is evidently equivalent to the dilaton model (\ref{obh:Vdil}) with 
\begin{equation}
{\cal F}(\Phi)={\omega(\Phi)\over 2}, \qquad 
{\cal G}(\Phi)={(\omega'(\Phi))^2\over 2 \xi(\Phi)}.\label{obh:FGP}
\end{equation} 
The equivalence of (\ref{obh:V0}) and (\ref{obh:eqV0}) can be also verified by
comparing the relevant sets of field equations.

The scalar-tensor PGG model (\ref{obh:V0}) can be straightforwardly
generalized in order to include the potential for the dilation field
$\Phi$ and a possible interaction with the matter field $\Psi$:
\begin{eqnarray}
L_{\rm tot} &=& V + L_{\rm mat}(\Psi, D\Psi, \Phi), \label{obh:Ltot}\\
V &=& -\,{1\over 2}\,\xi(\Phi)T_\alpha\wedge\ast T^\alpha 
-{1\over 2}\,\omega(\Phi) R^{\alpha\beta}\eta_{\alpha\beta} +
{\cal U}(\Phi)\,\eta.\label{obh:Vtot}
\end{eqnarray}
As we know, for standard matter in two dimensions (scalar, spinor,
Abelian and non--Abelian gauge fields), the spin current is zero,
$\tau_{\alpha\beta}=0$, and hence the ``second" field equation
(\ref{obh:TdP}), which results from varying the connection, remains the
same for the generalized model (\ref{obh:Vtot}).  This fact is the basis
of the equivalence of the scalar-tensor model (\ref{obh:Vtot}) and the
general dilaton type model (\ref{obh:Vdil}) with the same identifications
of the coefficient functions (\ref{obh:FGP}). The dilaton field potential
${\cal U}(\Phi)$ is taken from (\ref{obh:Vdil}), whereas
\begin{equation}
L_{\rm mat}=2{\cal J}(\Phi)\,F_B\wedge\ast F^B =
{\cal J}(\Phi)\,(F^B_{ij})^2\,\eta\label{obh:YM}
\end{equation}
represents a specific matter field $\Psi$, with $\Psi=A^B$. In general,
we may have a larger set of matter fields.

\subsection{Proof}\label{obh:Proof}

The explicit construction of the general solution for dilaton gravity,
(\ref{obh:dsdil}), (\ref{obh:dilh}), and (\ref{obh:dilnu}), is obtained by the
same machinery as that developed for two-di\-men\-sion\-al PGG.

Again, we have a Lagrangian which depends on the torsion square $t^2$
and the curvature scalar $R$. The gravitational field momenta
(\ref{obh:KP}) read
\begin{equation}
 P=\xi(\Phi),\qquad\kappa=-\,\omega(\Phi).\label{obh:KP1}
\end{equation}
The Lagrangian two--form (\ref{obh:Vtot}) can be transformed into the
corresponding Lagrangian density in (\ref{obh:calv}): ${\cal V} = -{1\over
  2}(\xi\,t^2 + \omega\,R) - {\cal U}$. Hence (\ref{obh:cv}) yields
$\widetilde{\cal V} = {1\over 2} \xi\,t^2 - {\cal U}$.

In the absence of matter, which can be enforced in (\ref{obh:YM}) by
putting ${\cal J}(\Phi) =0$, we immediately obtain the gravitational
field equations (\ref{obh:dpt2})-(\ref{obh:dk}) in coordinate and gauge
invariant form:
\begin{eqnarray}
d(\xi^{2}t^{2}) &=& 2\left({\cal U}(\Phi) - {1\over 2}\xi(\Phi)\,t^2\right)
d\,\kappa,\label{obh:dpt2dil}\\
d(\xi{\ast}T) &=& 2\,{\cal U}(\Phi)\,\eta,\label{obh:dptdil}\\
d\,\kappa &=& - \xi\, T.\label{obh:dkdil}
\end{eqnarray}
If a Yang--Mills field is present, its contribution manifests itself
in a simple ``deformation'' of the potential function ${\cal
  U}(\Phi)$, similar to the shift of the cosmological constant in
(\ref{obh:cosm}). Indeed, taking into account the Yang-Mills field
equation for (\ref{obh:YM}),
\begin{equation}
D({\cal J}(\Phi)\ast F_A) =d({\cal J}(\Phi)f_A) + 
c_{ABC} A^B\,{\cal J}(\Phi)f^C =0,
\end{equation}
where $f_A:=\ast F_A$, we straightforwardly obtain the first integral
\begin{equation}
f^Bf_B\,({\cal J}(\Phi))^2=:Q_0^2.
\end{equation}
With the help of this result, one can prove that the field equations
(\ref{obh:dpt2dil}),(\ref{obh:dptdil}) are formally the same, if the potential
${\cal U}(\Phi)$ is replaced by
\begin{equation}
{\cal U}(\Phi)\longrightarrow
{\cal U}(\Phi) + {1\over 2}Q_0^2/{\cal J}(\Phi).\label{obh:UU}
\end{equation}
In order to simplify the
notation, we will treat both cases simultaneously, considering the system 
(\ref{obh:dpt2dil})-(\ref{obh:dkdil}) with ${\cal U}(\Phi)$ properly defined. 

The integration of (\ref{obh:dpt2dil})-(\ref{obh:dkdil}) is straightforward. At
first, after substituting (\ref{obh:KP1}), we immediately obtain a linear
equation for $\xi^2 t^2$,
\begin{equation}
\left(\xi^2\,t^2\right)' = {\omega'\over \xi}\,\xi^2\,t^2
- 2\,{\cal U}\omega',\label{obh:alt2}
\end{equation}
where the prime denotes a derivative with respect to $\Phi$. Formal
integration yields
\begin{eqnarray}
-\,\xi^2\,t^2&=&2\,e^{\nu(\Phi)}\,\left(M_0 + \int\limits_{}^{\Phi}
d\varphi\,{\cal U}(\varphi)\,\omega'(\varphi)\,e^{-\nu(\varphi)}\right),
\label{obh:altsol}\\
\nu(\Phi)&=&\int\limits_{}^{\Phi}d\varphi\,{\omega'(\varphi)\over
\xi(\varphi)}.\label{obh:nusol}
\end{eqnarray}

The construction of the metric can be completed along the lines of our
general method. Namely, since $t^2\neq 0$, one can construct a
zweibein from the torsion one-form $T$ and its dual $\ast T$.
According to (\ref{obh:dkdil}), we may consider either $\kappa
=-\omega(\Phi)$ or the the field $\Phi$ itself as a first local
coordinate. The second (``time'') coordinate, say $\lambda$, is then
naturally associated with another leg of the zweibein,
\begin{equation}
\xi\ast\! T := B\,d\lambda,
\end{equation}
with some function $B=B(\lambda,\kappa(\Phi))$. Substituting the
latter into (\ref{obh:dptdil}) and taking into account the volume 2--form
$\eta={B\over \xi^2t^2}\,d\kappa\wedge d\lambda$, see (\ref{obh:vol}), we
obtain an equation for $B$:
\begin{equation}
{\partial\ln B\over\partial\kappa} = {2{\cal U}\over \xi^2t^2}.
\end{equation}
On integration
\begin{equation}
B=B_0(\lambda)\,\xi^2t^2\exp\left(\int {d\kappa\over\xi}\right)=
B_0(\lambda)\xi^2t^2\,e^{-\nu(\Phi)},
\end{equation}
and, again without loss of generality, one can put the function 
$B_0(\lambda)=1$. 

Accordingly, the metric finally reads:
\begin{equation}
g=-{d\kappa^2 \over \xi^2t^2} + \xi^2t^2 \exp\left(2\int 
{d\kappa\over \xi}\right)d\lambda^2 = 
\xi^2t^2\,e^{-2\nu(\Phi)}d\lambda^2 - {(\omega')^2\over\xi^2t^2}
\,d\Phi^2.\label{obh:metdil}
\end{equation}
Recalling the identifications (\ref{obh:FGP}), which establish the
relation between the dilaton and PGG, we arrive at the result
(\ref{obh:dsdil}) by putting $h(\Phi):=-\xi^2t^2/4$. Then the equations
(\ref{obh:altsol}) and (\ref{obh:nusol}) reduce to (\ref{obh:dilh}) and
(\ref{obh:dilnu}), respectively.

\subsection{Concluding remarks}

The general solution described in Sec.\ \ref{obh:sol} contains all the
exact black hole configurations reported earlier in the literature as
particular cases which correspond to specific choices of the
coefficient functions ${\cal F}(\Phi), {\cal G}(\Phi), {\cal U}(\Phi),
{\cal J}(\Phi)$, cf.\ \cite{cal,nappi1,nappi2,nappi3,russo}, e.g..

The new results are most useful for the investigation of the dynamical
picture of a gravitational collapse for nontrivial matter sources.  Of
particular interest is the case of a non-minimally coupled scalar
field which describes the semi-classical correction to the Lagrangian
due to Hawking radiation.

\subsection*{Acknowledgments}
We are grateful to Marc Toussaint for useful remarks and for reading
the manu\-script and to Ralph Metzler for help in drawing the figures.

\end{document}